\documentclass[a4paper]{jpconf}
\usepackage{graphicx}
\begin{document}
\title{(In)Direct Detection of Boosted Dark Matter}

\author{Kaustubh Agashe$^\dag$, Yanou Cui$^{\dag \flat}$ , Lina Necib$^\ddag$\footnote{Speaker at TAUP conference, Turin, Italy Sept 7-11, 2015}, Jesse Thaler$^\ddag$ }

\address{$^\dag$Maryland
Center for Fundamental Physics, University of Maryland,
College Park, MD 20742, USA \\
$^\flat$Perimeter Institute, 31 Caroline Street North Waterloo, Ontario N2L 2Y5, Canada\\
$^\ddag$Center for Theoretical Physics, Massachusetts Institute of Technology,
Cambridge, MA 02139, USA}

\ead{kagashe@umd.edu, ycui@perimeterinstitute.ca, lnecib@mit.edu, jthaler@mit.edu}

\begin{abstract}
We present a new multi-component dark matter model with a novel experimental signature that mimics neutral current interactions at neutrino detectors. In our model, the dark matter is composed of two particles, a heavier dominant component that annihilates to produce a boosted lighter component that we refer to as boosted dark matter. The lighter component is relativistic and scatters off electrons in neutrino experiments to produce Cherenkov light. This model combines the indirect detection of the dominant component with the direct detection of the boosted dark matter. Directionality can be used to distinguish the dark matter signal from the atmospheric neutrino background. We discuss the viable region of parameter space in current and future experiments.
\end{abstract}

\section{Introduction}

Despite overwhelming gravitational evidence for the existence of dark matter (DM), this non-baryonic matter has so far evaded all other means of observation.  It is therefore essential to investigate less conventional models for DM with non-traditional experimental signatures.
In these proceedings, we discuss a new DM model with novel signals at neutrino experiments. Our DM model involves two components, a dominant secluded GeV scale species $\psi_A$, that annihilates into a lighter species $\psi_B$ at regions of high DM density, and in particular the galactic center (GC). The $\psi_B$ particles are produced with a large boost, and we therefore term them boosted DM. Detection of $\psi_B$ occurs through the scattering $\psi_B e^- \rightarrow \psi_B e^-$ at neutrino experiments. The background to this process consists of electrons produced from charged current atmospheric neutrino reactions. Due to the kinematics of the signal, the scattered electron is emitted in the forward direction and a search cone can be used to optimize for signal significance and results in a detectable signal in current and future neutrino experiments.  A more detailed discussion is presented in Ref. \cite{Agashe:2014yua}.
 
\section{Two-component Dark Matter}
\label{sec:model}

We begin by outlining the particle content of our model. We assume that DM is comprised of two Dirac fermions: $\psi_A$ and $\psi_B$. We take $\psi_A$ to be the dominant DM component, which annihilates to $\psi_B$ via the dimension six operator
\begin{equation}
\mathcal{L}_{\rm{int}} = \frac{1}{\Lambda^2} \overline{\psi_A} \psi_B \overline{\psi_B} \psi_A.   
\end{equation}
The scale of the process $\Lambda \sim \mathcal{O} (100 ~\rm{GeV})$ is set by the requirement that the abundance of $\psi_A$ matches the observed DM abundance. The abundance of $\psi_A$ is controlled by the annihilation rate $\overline{\psi_A} \psi_A \rightarrow \overline{\psi_B} \psi_B$ which requires $m_A > m_B$ to be kinematically allowed.

The particle $\psi_B$ is charged under a dark force, mediated by a dark photon $\gamma'$, that kinematically mixes with the photon\footnote{More accurately, the dark $U(1)$ mixes with hypercharge.} through the interaction
\begin{equation}
 \mathcal{L} \supset - \frac{\epsilon}{2} F'_{\mu \nu} F^{\mu \nu},
\end{equation}
where $F^{\mu \nu}$ is the photon field and $F'^{\mu \nu}$ is the dark photon field. The coupling $\epsilon \sim \mathcal{O}(10^{-3})$ sets the scale of the mixing. The gauge coupling of this dark force is taken to be large yet perturbative $g'\sim \mathcal{O}(0.1)$. The processes that control the abundance of $\psi_B$ are the annihilation of $\psi_A$: $\overline{\psi_A} \psi_A  \rightarrow \overline{\psi_B} \psi_B $ and the $\psi_B$ annihilation to dark photons: $\overline{\psi_B} \psi_B \rightarrow \gamma' \gamma'$, where we have assumed $m_B > m_{\gamma'}$. The coupling of $\psi_B$ to the dark photon is taken such that the abundance of thermal $\psi_B$ is subdominant to that of $\psi_A$.

Throughout this paper, we use the benchmark\footnote{Since the publication of this work, this combination for $(m_{\gamma'}, \epsilon)$ has been excluded by the preliminary work of the NA48/2 collaboration \cite{Goudzovski:2014rwa}. A new benchmark can be found in the parameter space around this point.} 
\begin{equation}
\label{eq:benchmark}
 m_A = 20 ~{\rm{GeV}}, \qquad m_B= 200~{\rm{MeV}}, \qquad m_{\gamma'}= 20~{\rm{MeV}}, \qquad g' = 0.5, \qquad \epsilon =10^{-3},
\end{equation}
to quantify scattering cross sections.
In Sec. \ref{sec:results}, we explore the full range of parameter space, carefully taking into account experimental constraints, and discuss prospects for future discovery.

\section{Dark Matter Production Mechanism}
\label{sec:production}
In our model, while $\psi_A$ provides the dominant DM density, $\psi_B$ particles produced through the $\overline{\psi_A} \psi_A \rightarrow \overline{\psi_B} \psi_B$ process today provide a candidate for detection. Production of $\psi_B$ occurs predominantly in region of high DM ($\psi_A$) density, in particular the GC.
The flux of $\psi_B$ at the GC is
\begin{equation}
\label{eq:flux}
\frac{d \Phi_{\rm{GC}}}{d \Omega \, d E_B} = \frac{1}{4} \frac{r_{\rm{Sun}}}{4 \pi} \left( \frac{\rho_{\rm{local}}}{m_A}\right) ^2 J \,  \langle \sigma_{A\overline{A} \rightarrow B\overline{B}} v\rangle_{v \rightarrow 0} \frac{d N_B}{dE_B},
\end{equation}
where $r_{\rm{Sun}} = 8.33~$kpc is the distance from the galactic center to the Sun, $\rho_{\rm{local}} = 0.3 $ GeV/cm$^3$ is the local density of DM, and $\langle \sigma_{A\overline{A} \rightarrow B\overline{B}} v\rangle_{v \rightarrow 0}$ is the thermal cross section for the production of $\psi_B$, which we assume is dominated by the $s-$wave process. 
The spectrum of $\psi_B$ is given by
\begin{equation}
 \frac{dN_B}{dE_B} = 2 \delta(E_B - m_A),
\end{equation}
since every annihilation of $\psi_A$ produces two $\psi_B$ particles with energies $E_B = m_A$. Due to the mass hierarchy $m_A \approx \mathcal{O} (10)$ GeV  $ \gg m_B \approx \mathcal{O} (100) $ MeV, the $\psi_B$ particles are produced with a large boost, and we therefore refer to $\psi_B$ as boosted DM. Subsequently, $\psi_B$ can be detected at neutrino experiments. 

The DM density is incorporated in Eq. (\ref{eq:flux}) through the $J$-factor of the GC which is defined by
\begin{equation}
\label{eq:jfactor}
 J = \int_{\rm{l.o.s}} \frac{d s }{r_{\rm{Sun}}} \left( \frac{\rho(r(s,\theta))}{\rho_{\rm{local}}} \right)^2, 
\end{equation}
where the integral is along the line of sight, with $s$ the distance from an earth-located observer to a point in the DM halo. The coordinate $r$ is the distance from the center of the halo and is related to $s$ as
\begin{equation}
 r(s, \theta) = (r_{\rm{Sun}}^2 + s^2 - 2 r_{\rm{Sun}} s \cos{\theta})^{1/2}.
\end{equation}
The angle $\theta$ is the angle between a point in the halo and the center of the galaxy from the point of view of an observer on earth. For simplicity, we assume an NFW profile for the DM density distribution $\rho(r(s,\theta))$, and use the numerical code of Ref. \cite{Cirelli:2010xx} to compute the $J$-factor as a function of the angle $\theta$. 
Having found the boosted DM $\psi_B$ flux, we use the differential cross section of the process $\psi_B e^- \rightarrow \psi_B e^-$ to infer the expected number of signal events.

\section{Kinematics of Boosted DM Detection}
\label{sec:detection}

\begin{figure}[t]
\begin{center}
\includegraphics[width=13pc, trim = 0 0 0 0]{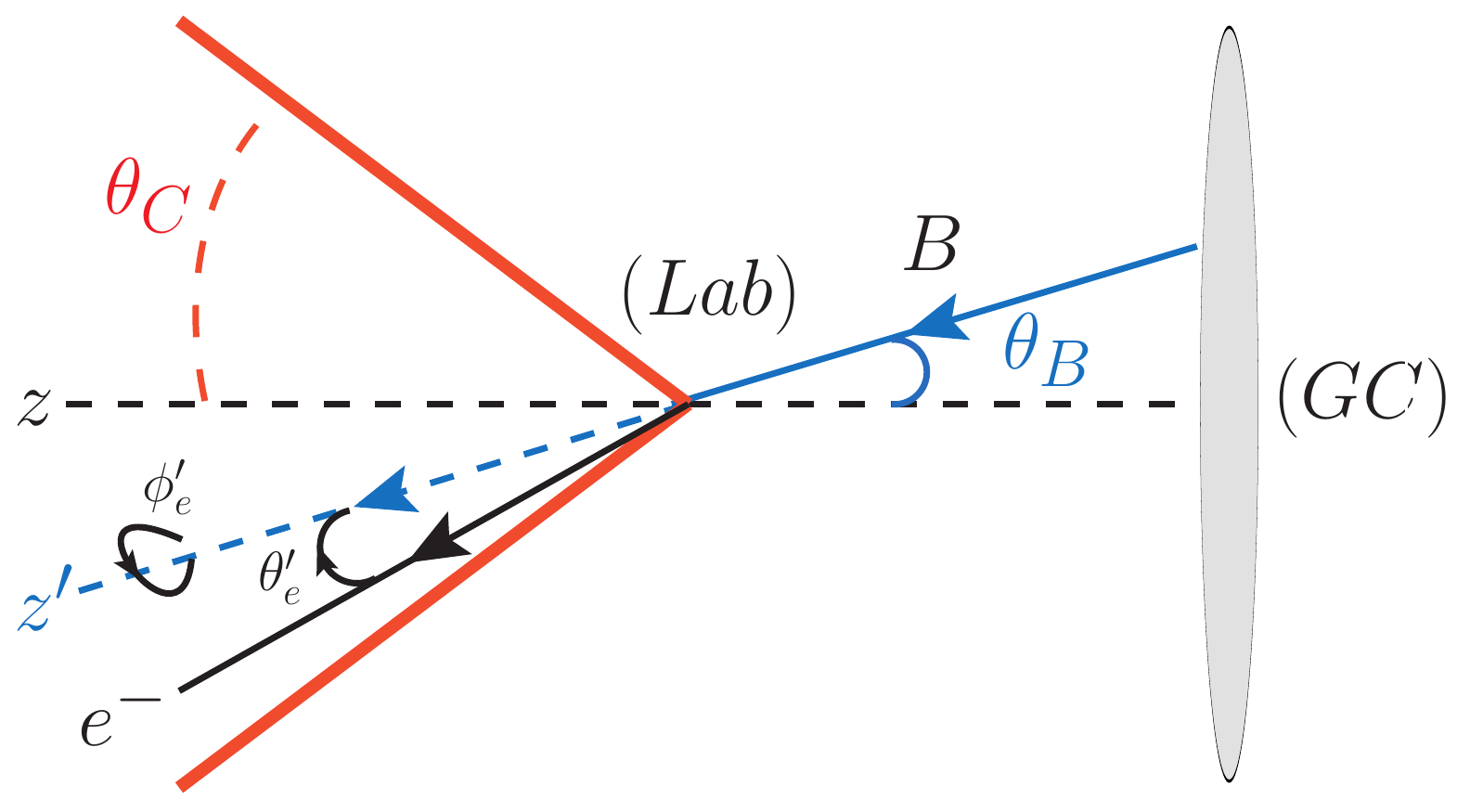}\hspace{2pc}%
\caption{\label{fig:cone} Geometry of a search cone for incoming $\psi_B$ particles originating at the GC and scattering off electrons at neutrino experiments. }
\end{center}
\end{figure}

Proceeding from the production mechanism of the boosted DM component $\psi_B$ to its detection, we set up a framework to optimize for the observation of $\psi_B$ through electron scattering in neutrino detectors. Due to the kinematics of $\psi_B e^- \rightarrow \psi_B e^-$ scattering, the electron is emitted in the forward direction and thus correlates with the direction of the incoming $\psi_B$, as schematically shown in Fig. \ref{fig:cone}. Since the GC is the densest region of DM, we search for events where the emitted electron, and by extension the incoming $\psi_B$, is within an angle $\theta_C$ of the GC.
 The angle $\theta_C$ can be optimized for maximal discrimination between the boosted DM signal and the background by studying the analytical form of the scattering cross section. This involves the kinematics of the $\psi_B$-electron scattering, and the convolution of the electron angular distribution with the initial DM distribution in the GC. We discuss each in turn.

The differential cross section for the $\psi_B e^- \rightarrow \psi_B e^-$ process is
\begin{equation}
\label{eq:diffBeBe}
\frac{d\sigma_{B e^- \rightarrow B e^-}}{dt }= \frac{1}{8 \pi}  \frac{ (\epsilon e g')^2}{(t - m_{\gamma'}^2)^2} \frac{8 E_B^2 m_e^2 + t(t+ 2s)}{\lambda(s,m_e^2, m_B^2)},
\end{equation}
where $\lambda (x,y,z)= x^2 + y^2 + z^2 - 2 x y - 2 xz - 2 yz$, $s = m_B^2 + m_e^2 + 2 E_B m_e$, $t = q^2 =  2 m_e (m_e - E_e)$.  The energy of $\psi_B$ is $m_A$, since $\psi_B$ is produced from the annihilation of $\psi_A$ at rest. To obtain the full cross section, Eq. (\ref{eq:diffBeBe}) must be integrated over the valid range of electron energies. The lower bound of integration is given by
$E_e^{\rm{min}} = E_e^{\rm thresh},$
as the energy of the outgoing electron has to be above the experimental threshold. The maximum scattered electron energy allowed by kinematics is
\begin{equation}
E_e^{\rm{max}} = m_e \frac{(E_B + m_e)^2 + E_B^2 - m_B^2}{(E_B + m_e)^2 - E_B^2 + m_B^2}. \label{eq:emax}
\end{equation}

For the benchmark values of Eq. (\ref{eq:benchmark}) and a minimum allowed energy of the electron  $E_e^{\rm{min}} = 100$ MeV, we find
\begin{equation}
\label{eq:typicalsignalxsec}
\sigma_{B e^- \rightarrow B e^-}  = 1.2 \times 10^{-33}~{\rm{cm}}^2 \left(\frac{\epsilon}{10^{-3}} \right)^2 \left(\frac{g'}{0.5} \right)^2 \left(\frac{20~{\rm{MeV}}}{m_{\gamma'}} \right)^2.
\end{equation}

Convolving the NFW distribution of $\psi_B$ with the angular distribution of the scattered electrons, we find the number of expected signal events in which the emitted electron is within an angle $\theta_C$ of the GC
\begin{eqnarray}
N^{\theta_C}_{\rm{signal}}&=& \Delta T \, N_{\rm{target}} \, \left(\Phi_{\rm{GC}} \otimes \sigma_{B e^- \rightarrow B e^-})\right) \Big|_{\theta_C} \nonumber \\
&=& \frac{1}{2} \Delta T \, \frac{10 \, \rho_{\rm{Water/Ice}} V_{\rm{exp}}}{m_{\rm{H_2O}}} \frac{r_{\rm{Sun}}}{4 \pi} \left( \frac{\rho_{\rm{local}}}{m_A}\right)^2 \langle \sigma_{A \overline{A} \rightarrow B \overline{B} } v\rangle_{v \rightarrow 0}  \\\nonumber
&&~\times  \int_0^{2 \pi} \frac{d \phi'_e}{ 2 \pi}  \int^{\theta'_{\rm max}}_{\theta'_{\rm min}} d \theta'_e \, \sin \theta'_e \, \frac{d \sigma_{B e^- \rightarrow B e^-}}{d \cos\theta'_e} \int_0^{\pi/2} d \theta_B \sin \theta_B  \, 2\pi J(\theta_{B}) \Theta(\theta_C-\theta_e). \label{eq:signalconvolve}
\end{eqnarray}
Here $\Delta T$ is the exposure time, $N_{\rm{target}} $ is the number of target electrons, and $\Phi_{\rm{GC}}$ is the $\psi_B$ flux. In order to find the number of targets per experiment, we use the density of the material (in this case water or ice) $\rho_{\rm{Water/Ice}}$, the volume of the experiment $V_{\rm{exp}}$, and the mass of a water molecule $m_{\rm{H_2O}}$. We parameterize the direction of the emitted electron by an azimuthal angle $\phi_e'$ and a polar angle $\theta_e'$. As shown in Fig. \ref{fig:cone}, the detection geometry respects a cylindrical symmetry, while the angle $\theta_e'$ is constrained by the minimum allowed energy of the electron (experimental resolution) and a maximum energy determined by both the kinematics of the scattering and the experimental setup that might divide events into Sub-GeV and Multi-GeV events, according the electron energy.

For the benchmark values given in Eq. (\ref{eq:benchmark}), the expected number of signal events within an angle $\theta_C = 10^\circ$ of the GC is
\begin{equation} \label{eq:signal}
\frac{N^{10^\circ}_{\rm{signal}}}{\Delta T} = 25.1 ~{\rm{year}}^{-1}.
\end{equation}

\section{Detection Prospects for Current and Future Experiments}
\label{sec:results}

In this section, we discuss the detection limits of boosted DM in Super-K \cite{Ashie:2005ik} and its upgrade Hyper-K \cite{Kearns:2013lea}, as well as the IceCube extensions PINGU \cite{Aartsen:2014oha} and MICA  \cite{MICA}. Despite its large volume, IceCube's energy threshold is too high for the considered DM scale  \cite{Abbasi:2011eq}.
At the energy scale of $\psi_B$, the largest background to $\psi_B e^- \rightarrow \psi_B e^-$ is atmospheric neutrinos. They interact through the charged current processes $\nu _e n \rightarrow e^- p$ and $\overline{\nu_e} p \rightarrow n e^+$, producing electrons of similar energies as signal events. A few features can help discriminate between signal and background:
\begin{itemize}
\item Directionality: The atmospheric neutrinos have an isotropic flux while boosted DM is focused at the direction of largest $J$-factor, which in this case is the GC. 
\item No muon events: While atmospheric neutrinos come in different flavors, and thus produce both electron and muon events, signal events produce solely electrons.
\item Gadolinium:  Experiments such as Super-K are currently investigating using Gadolinium to improve detection efficiency \cite{Beacom:2003nk}. Gadolinium has the highest neutron capture cross section and can thus discriminate with high efficiency against the $\overline{\nu_e} p \rightarrow n e^+$ background events. 
\end{itemize}

Extrapolating from the atmospheric neutrino data of Super-K \cite{Wendell:2010md}, we derive the expected number of background events at other experiments. Since the signal is analyzed within an angle $\theta_C$ of the GC, we only look at the fraction of background events that fall into the same solid angle
\begin{equation}
N_{\rm{bkgd}}^{\theta_C} = \frac{1 - \cos \theta_C}{2} N_{\rm{bkgd}}^{\rm{all sky}}.
\end{equation}

\begin{figure}[t]
\begin{center}
\includegraphics[width=13pc]{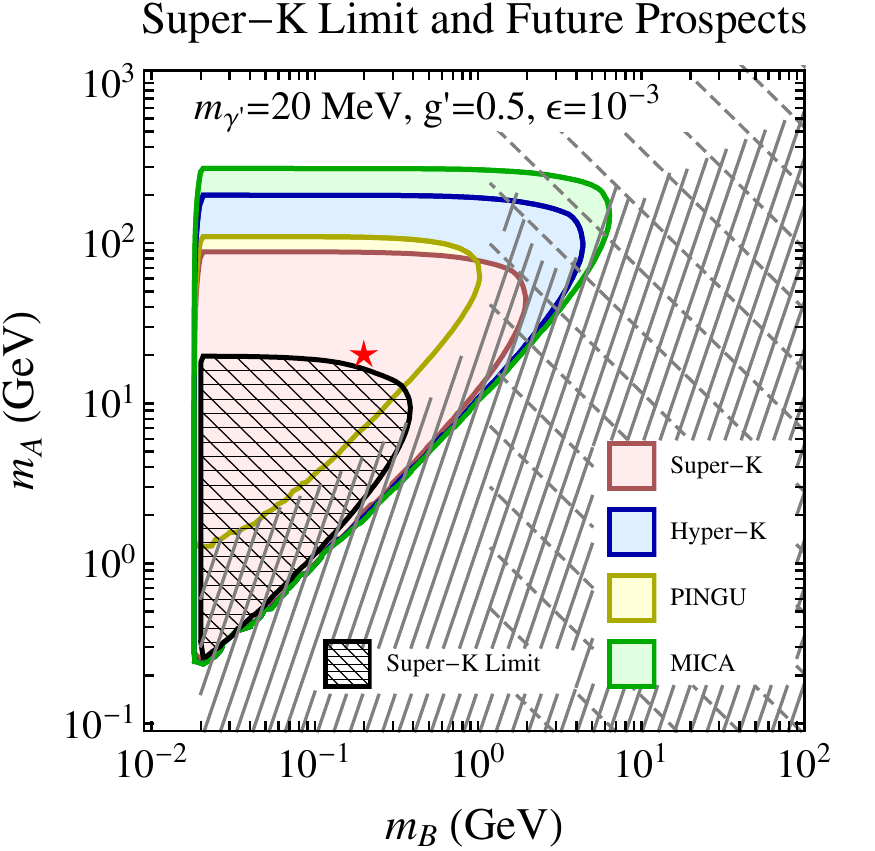}\hspace{2pc}%
\caption{\label{result} Signal significance at Super-K, Hyper-K, PINGU and MICA as a function of the parameter space $m_A - m_B$, for fixed $m_{\gamma'}, g' $ and $\epsilon$ with the benchmark values of Eq. (\ref{eq:benchmark}). Regions shown are $2 \sigma$ reaches for 10 years of data. }
\end{center}
\end{figure}

We define the signal significance as $
 \rm{Signal}/\sqrt{Bkg},
$
and use it as a measure to optimize for the opening angle of the search cone. We find that $\theta_C = \rm{max} (10^\circ, \theta_e^{\rm{res}})$, where $\theta_e^{\rm{res}}$ is the experimental angular resolution. In the case of PINGU and MICA , it exceeds $10^\circ$. 
For $\theta_C =10^\circ$, as relevant for Super-K, we have
\begin{equation}
\frac{N_{\rm{bkgd}}^{10^\circ}}{\Delta T} = 5.85~\rm{year}^{-1}. 
\end{equation}
Within $\theta_C = 10^\circ$, the number of background events is subdominant to the number of expected signal events given by Eq. (\ref{eq:signal}). The resulting significance as a function of $m_A $ and $m_B$ is shown in Fig. \ref{result}. Already, without the use of directionality, the hatched black region of parameter space is ruled out by Super-K.  A dedicated analysis using the suggested search cone would be able to extend its detection reach to encompass our benchmark values of Eq. (\ref{eq:benchmark}).

\section{Conclusions}
In these proceedings, we presented an example of a DM model that provides a new experimental signature. Starting from a multi-component model, we looked into the annihilation of a DM component resulting in a boosted DM particle. The boosted DM component, if coupled to the Standard Model through a dark photon, can scatter off electrons at neutrino experiments and produce detectable Cherenkov light. The signal can be distinguished from background events using the angular information of emitted electrons, as signal events are in the forward direction and point to the GC. Furthermore, unlike background events, mainly produced from atmospheric neutrinos, the signal electron events are not accompanied by muon events. Boosted DM is therefore an exciting new paradigm that opens the door to new experimental possibilities for DM detection.

\section*{References}
\bibliography{biblio}{}

\end{document}